# NEMD modeling of nanoscale hydrodynamics of clay-water system at elevated temperature

Zhe Zhang, Xiaoyu Song

*Engineering School of Sustainable Infrastructure and Environment, University of Florida, Gainesville,Florida USA 32611*

**Abstract**

The engineering problems involving clay are multiscale and multiphysics by nature. The nanoscale hydrodynamics of clay at elevated temperature is essential in developing a physics-based multiscale model for clay. The nonequilibrium molecular dynamics (NEMD) is a useful tool to study the nanoscale hydrodynamics of clay. This article presents an NEMD modeling of hydrodynamics of clay nanopores at elevated temperatures. Water flow confined in pyrophyllite and montmorillonite clay nanopores is investigated. The nonequilibrium state is maintained by uniformly exerting an external force on each water molecule. The NEMD simulations have provided a molecular-scale perspective of temperature effect on clay-water density, water flow velocity, shear viscosity, clay-water slip length, hydraulic conductivity, and clay-water friction coefficient. The numerical results have shown a strong temperature dependence of fluid flow velocity, shear viscosity, clay-water slip length, and hydraulic conductivity at the nanoscale. We have validated the applicability of cubic law in determining hydraulic conductivity at the nanopore scale under at elevated temperature. It is found from our numerical results that slip clay-water boundary condition is an essential factor in properly determining nanoscale fluid flow velocity. By numerical examples, we also study the impact of nanopore size and clay layer thickness on the hydrodynamics of the clay-water system.

## 1. Introduction

Hydrodynamics describes flow characteristics of a fluid in motion including density, velocity, viscosity, conductivity, and friction coefficient as functions of time, space, and/or temperature [1]. The knowledge of nanoscale hydrodynamics of clay under non-isothermal conditions is essential for the understanding of natural processes and engineering applications of clay, such as groundwater hydrology, oil/gas exploitation, and geological disposal of radioactive waste [2, 3, 4, 5]. Laboratory tests have been conducted to understand the hydrodynamic behavior of clay under non-isothermal conditions at the continuum scale (e.g., [6, 7, 8, 9]). For instance, Romero et al. [6] presented an experimental study focusing on the temperature impact on hydraulic properties of kaolinitic-illitic clay related to water retention and permeability. Villar and Lloret [7] reported the influence of temperature on hydro-mechanical properties of compacted bentonite from laboratory tests. They concluded that the decrease in water viscosity with temperature attributes to the increase of hydraulic conductivity. Tang and Cui [8] experimentally investigated thermal effects on hydraulic-mechanical behaviors of a swelling clay using the vapor equilibrium technique and found that the retention curve shifts downward with increasing temperature. Sun et al. [9] performed vapor transfer experiments on bentonite in a wide range of temperatures and found a systematic loss of water retention capacity of bentonite at high temperature. However, it is difficult to study the hydrodynamics of clay at

*Email address:* xysong@ufl.edu (Xiaoyu Song)

the molecular level through laboratory tests [10]. Indeed, the hydrodynamics of clay is multiscale and multiphysics by nature in that clay is a natural nanomaterial and the transport at a larger scale is controlled by the transport through the smallest pore scale [11]. Multi- scale modeling is a useful tool to study the hydrodynamics of clay at different temperatures. However, one major challenge in multiscale modeling of materials including clay [12, 13] is the faithful understanding of physical models at different space and time scales. Thus, to develop a physics-based multiscale model for fluid flow in clay that is a natural nanomaterial, we need to understand the nanoscale hydrodynamics of clay. In this article we investigate the nanoscale hydrodynamics of clay at elevated temperatures through non-equilibrium molecular dynamics modeling.

When water flows in nanopores, the influence of the liquid-solid wall interaction becomes more pronounced compared to a macro-pore of large volume. The induced inhomogeneity results in the layered water molecules close to the solid surface. Due to this inhomogeneity, traditional fluid dynamics models like the Navier-Stokes equation become invalid. For this reason, the evaluation of fluid-solid interaction requires an appropriate frictional/slip boundary condition. A breakthrough is the microscopic expressions for the interfacial mechanical and thermal slip transport coefficients [14]. In general, understanding hydrodynamic phenomena at the nanoscale requires fundamentally different mathematical modeling and numerical approaches than those in macroscopic frameworks. Laboratory tests have difficulty in revealing the mechanism of water transport in nanoscale networks. They suffer from a significant issue, the inability to provide fundamental insight into the adsorption mechanism near the clay-water interface at the molecular level. To this end, molecular simulations have contributed to study the hydraulic behavior and structure of clay minerals at the nanoscale [10]. Molecular dynamics (MD) is a computer simulation technique used to compute the movement of particles by solving Newton's equations of motion. MD has been applied to investigate the soil-water adsorption, soil-water contact angle structural property, and water diffusion in soils [15, 16, 17, 18, 19].

On the atomic scale, there are three main forces in clay, namely van der Waals forces, ionic bonds, and covalent bonds, which are arranged in the order of weak to strong. Covalent bonds are widely found in tetrahedral and octahedral sheets in clay. Due to these strong bonds, the structural stability of clay is retained when clay interacts with water [20]. An MD simulation requires appropriate potential functions, also known as force fields, to compute the forces between atoms. In geomechanics, several force fields have been formulated to investigate mechanical and physicochemical properties of clay, such as CHARMM [21], universal force field [22], valence force field [23], and CLAYFF [24]. For instance, Bains et al. [25] performed Mont Carlo and molecular dynamics simulations with the universal force field to investigate the mechanism of inhibition of clay-swelling. Their results indicate that entropy is the driving force for the sorption of the simpler organic molecules inside the clay layers [25]. Katti et al. [26] proposed a linear displacement-stress relationship for pyrophyllite interlayer through MD simulations with the CHARMM force field. Song and Wang [13] examined the temperature effect on the capillary force and capillary pressure of unsaturated clay using the CHARMM force field. Their numerical results showed that temperature increase generally decrease the capillary force and matric suction [13]. Ebrahimi et al. [27] simulated adsorption of interlayer water in Wyoming Na-montmorillonite using CLAYFF force field and derived the full elastic tensor calculations for this material over a wide range of hydration conditions.

Nonequilibrium molecular dynamics (NEMD) is a viable tool to investigate the nanoscale hydrodynamics of clay-water under non-isothermal conditions. NEMD simulation deals with properties of fluid under nonequilibrium conditions. Hence, it allows investigating the hydrodynamics of fluid under confinement. In NEMD simulations, the transport behavior of a fluid is characterized by measuring macroscopic steady-state response to an externally applied field [28]. By imposing an external force on water molecules, Gosling et al. [29] obtained the velocity profile of a Lennard-Jones fluid and further related its velocity to shear viscosity. The advantage of applying an external force instead of actual pressure to drive the fluid flow is that the confined water can remain longitudinally homogeneous under a constant external field [30]. This is a crucial factor for the planar Poiseuille flow through a narrow pore [31]. This approach has been extensively used in other studies to simulate the transport behaviors of nanoconfined water flow [32, 33, 34].

Hydrodynamic behavior and structure of water confined in clay interlayer depend on various factors including clay-water interaction strength, temperature, pore size, and external force [35, 36, 4, 37, 38, 39, 40, 41, 42]. Among these factors, temperature variations could play a key role in localized failure of variably saturated soils (e.g., [43, 44, 45]). Furthermore, temperature increase in soils could arise from climate change (e.g., global warming) or extreme events (e.g., earthquake and wildfire) [20]. Numerous studies of hydrodynamics of clay-water systems are based on experimental tests or numerical simulations conducted at ambient temperature. The knowledge of temperature impact on the nanoscale hydrodynamics of clay-water systems, to the best of our knowledge, is still limited. In this article, we investigate the temperature effect on nanoscale hydrodynamics of clays through NEMD. Section 2 presents the molecular structures of the materials and the technical details of the modeling method. Section 3 reports the numerical results of nanoscale dynamics at elevated temperature from NEMD simulations. Section 4 discusses the impact of temperature on soil-water absorption and clay-water friction and the effect of pore size on nanoscale hydrodynamics, followed by a closure in Section 5.

## 2. Materials and modeling methods

### 2.1. Molecular structures of clay and water

In this work, we study two types of clay minerals, pyrophyllite and Na-montmorillonite with chemical formulas $Al_2[Si_4O_{10}](OH)_2$ and $Na_{0.333}[Si_4O_8][Al_{1.667}Mg_{0.333}O_2(OH)_2]$, respectively. Figure 1 (a), (b), and (c) present the molecular structure of pyrophyllite, Na-montmorillonite, and water, respectively. The difference between the two clay minerals is that there is a negative charge in the octahedral sheet of montmorillonite due to the substitution of $Al^{3+}$ by $Mg^{2+}$. Pyrophyllite has no cation exchange capacity because no substitution takes place. The cation exchange capacity of Na-montmorillonite used in this work is 90 mequiv/100g. This type of montmorillonite has been widely used in the literature (e.g., [46, 47, 48]). The coordinates for the unit-cell of pyrophyllite can be found from Skipper et al. [49].

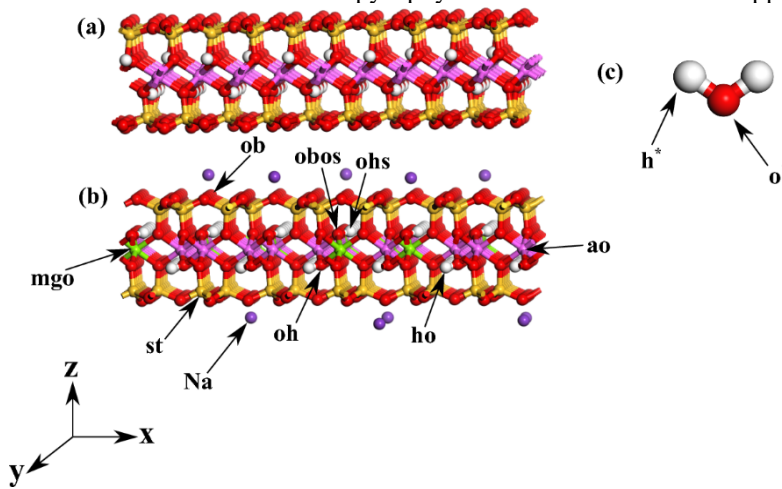

Figure 1: Molecular structures of (a) pyrophyllite, (b) Na-montmorillonite, and (c) water.

### 2.2. Force field for clay and water

In this study, the CLAYFF force field is adopted. CLAYFF is a general force field suitable for simulating hydration of clay minerals based on non-bonded interactions [50]. It has been proved that this force field can correctly predict the structural and thermodynamical properties of various clay minerals, as well as interlayer water and exchangeable cations (e.g., [51, 52, 53, 54]). Compared to other force fields based on quantum chemical calculations [23, 55, 56], CLAYFF incorporates some structural modifications from experiments. A major feature of CLAYFF is that it treats the great majority of the interactions between atoms as non-bonds. Thus, we can simulate various phases of larger clay-water systems with fewer parameters (i.e., fewer bonds and angles) [24]. In this research we focus on hydrodynamics of clay-water systems accounting for solid-liquid interactions. Thus we use CLAYFF to calculate the potential energy of clay-water systems.

In CLAYFF, each atom is assigned charge and van der Waals parameters and O-H groups are assigned bond stretch parameters. The total potential energy consists of nonbonded and bonded interactions (i.e., van der Waals and Coulombic energy, bond stretch and angle bend energies). The van der Waals energy can be written as

$$E_v = \sum_{i \neq j} D_{ij} \left[ \left( \frac{R_{ij}}{r_{ij}} \right)^{12} - 2 \left( \frac{R_{ij}}{r_{ij}} \right)^6 \right], \tag{1}$$

where $r_{ij}$ is the distance between atoms $i$ and $j$, and $D_{ij}$ and $R_{ij}$ are respectively the depth of the potential well and the distance at which the particle-particle potential energy vanishes between atoms $i$ and $j$. $D_{ij}$ and $R_{ij}$ can be calculated by

$$R_{ij} = \frac{1}{2}(R_i + R_j), \qquad D_{ij} = \sqrt{D_i D_j}, \tag{2}$$

where $D_i$, $R_i$ and $D_j$, $R_j$ are the input parameters for atoms $i$ and $j$, respectively. The Coulombic energy reads

$$E_c = \frac{1}{4\pi\epsilon_0} \sum_{i \neq j} \frac{q_i q_j}{r_{ij}}, \tag{3}$$

where $q_i$ and $q_j$ are the charges of atom $i$ and $j$, respectively, and $\epsilon_0$ is the vacuum permittivity. Nonbonded parameters in CLAYFF are compatible with the SPC/E (extended simple point charge) water model [57]. The H-O bond length and H-O-H angle in equilibrium are 1.0 Å and 109.47°, respectively. Nonbonded (e.g., Long-range electrostatic) interactions are cut off at 10 Å Long-range electrostatic interactions beyond 10 Å are evaluated using the particle-particle particle-mesh solver of accuracy 0.0001 [58]. Table 1 summarizes the values of the non-bonded parameters for the clay-water molecular model used in this study.

Table 1: Nonbonded parameters for the clay-water molecular model adopted in this study

| Species | Symbol | q (e) | D (kcal/mol) | R (Å) |
|---|---|---|---|---|
| water hydrogen | h* | 0.4238 | 0 | 0 |
| water oxygen | o* | -0.8476 | 0.1553 | 3.166 |
| hydroxyl oxygen | oh | -0.9500 | 0.1554 | 3.5532 |
| hydroxyl hydrogen | ho | 0.425 | 0 | 0 |
| bridging oxygen | ob | -1.0500 | 0.1554 | 3.5532 |
| bridging oxygen with octahedral substitution | obos | -1.0808 | 0.1554 | 3.5532 |
| hydroxyl oxygen with substitution | ohs | -1.0808 | 0.1554 | 3.5532 |
| tetrahedral silicon | st | 2.1000 | 1.8408e-6 | 3.7064 |
| octahedral aluminum | ao | 1.5750 | 1.3298e-6 | 4.7963 |
| octahedral magnesium | mgo | 1.3600 | 9.0298e-7 | 5.9090 |
| aqueous sodium ion | Na | 1.0 | 0.1301 | 2.6378 |

*2.3. Model setup*

Figure 2 presents the configuration of montmorillonite-water system with an 8 nm-wide pore. Each clay layer consists of 10 × 6 × 1 unit cells, which corresponds to 51.918 Å × 54.0918 Å in the x-y plane. Two clay layers are placed in parallel to form a pore space. The simulation boundary is periodic in all directions. The pore width should be large enough since the continuum approximation may be invalid in nanoscale confined systems. As stated in Travis et al. [59] the simulated velocity profile could be significantly different from the one predicted by the Navier-Stokes equation if the pore width is smaller than 5.1 water molecular diameters. Thus, in this study the pore width in MD models is taken as 8 nm. The number of water molecules $n_w$ introduced into the pore is determined such that the initial bulk density $\rho$ corresponds to 1 g/cm$^3$

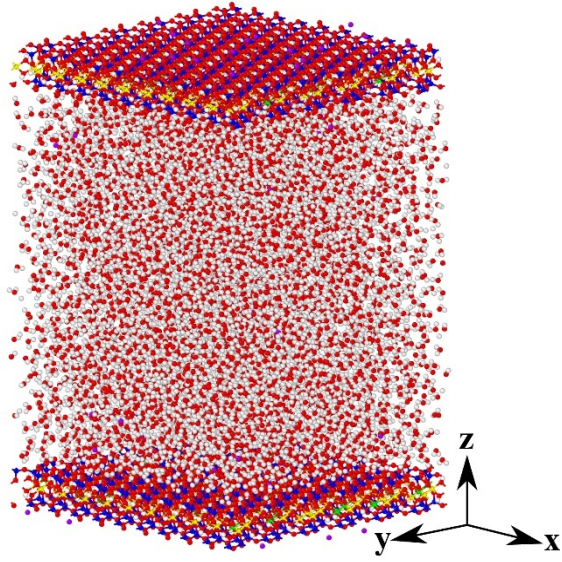

Figure 2: Snapshot of clay-water system during simulation.

$$n_w = \frac{\rho V N_A}{M_w},\quad (4)$$

where $V$ is the pore volume, $N_A$ is the Avogadro constant ($6.02 \times 10^{23}$ mol$^{-1}$), and $M_w$ is the molar mass of water. All MD simulations are performed with LAMMPS [60], a large-scaleatomic/molecular massively parallel simulator. SHAKE algorithm is employed to integratethe equations of motion for the rigid SPC/E water molecules [61]. This algorithm appliesconstraints to the bond length and bend angle of a water molecule so that water molecule istreated as a rigid body. The clay-water system was equilibrated in the canonical NVT ensemble.We first conducted the simulation in 200 fs (1 fs = $1 \times 10^{-15}$ s) with a time step 0.01 fs tominimize the energy of the clay-water system. Then we ran the simulation in 2 ns with atime step 0.5 fs for the system to reach an equilibrium state. Figure 3 plots the time-averagedpotential energy at several elevated temperatures. It indicates that 2 ns is appropriate time toobtain the steady state. The temperature is maintained by a Nose-Hoover thermostat [62]. Inthis study, we run simulations in a wide range of temperatures, from 300 K to 340 K.

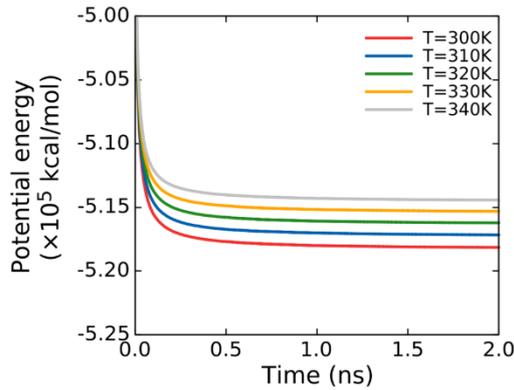

Figure 3: Time variations of potential energy at elevated temperatures.

### 2.4. Nonequilibrium molecular dynamics

A few nonequilibrium molecular dynamics methods have been developed to model fluid flow confined in nanopores. One is the external field nonequilibrium molecular dynamics [59]. The second technique is the dual control volume molecular dynamics where two rigid mobile walls are added at both ends acting like a piston [63]. We use the former method to model water flow because the external field NEMD method yields identical transport coefficients for many systems [64]. Besides, the external field NEMD technique is convenient to implement and is computationally efficient while preserving reliable results.

NEMD requires a large perturbation strength at the macroscale to produce a detectable microscale response. A response greater than the surrounding statistical noise of the system has to be produced. For this reason, we apply a relatively large external force on each water molecule. This is commonly used in NEMD simulations to reduce the thermal noise and to speed up the simulation process [65]. In order to be consistent with the former equilibrium MD simulation, we use the same timestep 0.5 fs for NEMD. We check the stable state of nonequilibrium by monitoring water streaming velocity over time. Figure 4 compares velocity profiles in different time periods. It is assumed that the NEMD simulation has reached a steady state at 4.5 ns since flow velocity profiles are almost identical after that.

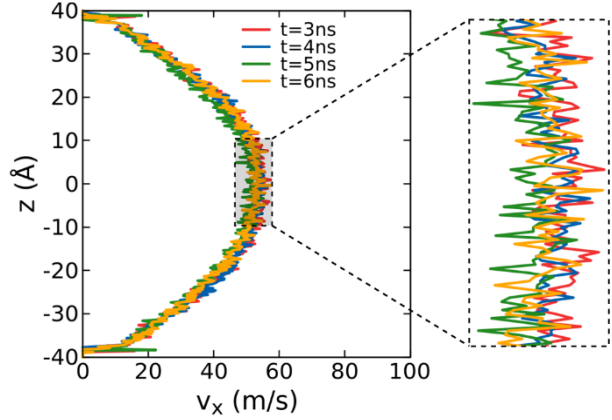

Figure 4: Time variations of velocity profiles.

## 3. Numerical results

We conduct NEMD simulations to study the temperature effect on nanoscale hydrodynamics of clay-water system. Following the external field NEMD technique, we exert a constant force along x-direction, $F_x = 0.0005$ kcal/(mol Å), to each water particle. Typically the fluid temperature is calculated through the kinetic energy by summing up the squares of velocities of all atoms. It is noted that in NEMD an extra velocity component from the center-of-mass movement of the whole water group is also involved. To remove the contribution of center-of-mass movement when calculating fluid temperature, we couple the thermostat only to the degree of freedom of water in y-direction. The relationship between applied force $F_x$ and pressure difference $\Delta P$ along flow direction is described as:

$$\Delta P = \frac{F_x N}{A_{yz}}, \tag{5}$$

where $N$ and $A_{yz}$ denote the total number of water atoms and the section area perpendicular to the flow direction, respectively [66].

In NEMD, density and velocity profiles are usually calculated by dividing the pore volume into several "bins" and averaging the properties of all molecules in each bin over time. By this means, we can spatially quantify the distribution and variations of density and velocity along z-direction. As shown in Figure 5, each bin represents a water layer of thickness $\Delta z$. For instance, the center-of-mass (COM) velocity of all water molecules in the bin is expressed as

$$v = \frac{\sum m_i v_{i,x}}{\sum m_i}, \tag{6}$$

where $v_{i,x}$ is the x-component velocity of particle $i$. The total NEMD simulation time is 6 ns for all cases, with a time step of 0.5 fs. All results are determined from the last 3 ns.

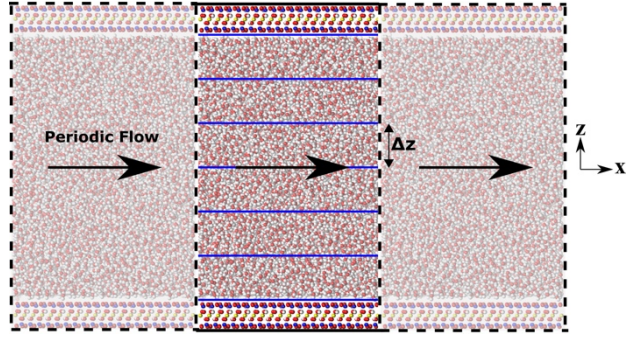

Figure 5: Schematic of periodic flow boundary and "binning" method.

*3.1. Confined water density*

Water density profile describes the distribution of water in clay nanopores. Figure 6 shows water density profiles in an 8 nm-wide clay pore at different temperatures. The density profile is symmetric with respect to $z = 0$. Water far away from the clay surface is stored in a bulk state that the bulk density almost keeps constant with the expected continuum value (1.03±0.006 g/cm$^3$). This agreement reinforces the validity and accuracy of our NEMD simulations. However, intense density fluctuations take place in the vicinity of clay-water interface caused by the strong intermolecular affinities between clay and water. The higher peak density, more tightly water molecules are adsorbed to clay surface. Table 2 summarizes the temperature effect on peak water density. Higher temperature results in greater kinematic energy of water flow, which further weakens the transport barrier due to clay adsorption. In pyrophyllite nanopore, the peak density decreases from 2.138 to 2.091 g/cm$^3$ as temperature increases by 40 K. Comparing the peak density between two clay-water systems, we find that Na-montmorillonite gives larger peak density than pyrophyllite (e.g., 2.319 g/cm$^3$ vs 2.138 g/cm$^3$ at 300 K). The reason could be that in situ sodium cations near montmorillonite surface re-orientates neighboring water molecules through ionic bond between cation and oxygen ion of water [20]. From simulation results, we find the density variations extend to as much as 13 Å from the clay surface. Beyond this distance, water density becomes independent of the clay surface. The length of density oscillations agrees well with the reported length 12 Å in [40].

Table 2: Temperature dependence of peak density.

| Temperature (K) | Peak density (g/cm$^3$) | |
|---|---|---|
| | Pyrophyllite | Montmorillonite |
| 300 | 2.138 | 2.319 |
| 310 | 2.130 | 2.306 |
| 320 | 2.124 | 2.302 |
| 330 | 2.110 | 2.296 |
| 340 | 2.091 | 2.289 |

*3.2. Velocity profile and shear viscosity*

The water density and viscosity are assumed to be constant in the center ($z = 0$) of the nanopore. This assumption is reasonable in that water at the center is far enough from the clay surface. If ignoring the slip boundary condition, the Navier-Stoke equation predicts a parabolic steady-state velocity profile for incompressible laminar flow.

$$v_x(z) = -\frac{nF_x}{2\eta}\left(z^2 - \frac{w^2}{4}\right), \qquad (7)$$

where $v_x(z)$ is water flow velocity in x-direction at $z$, $n$ is the number density of water, $F_x$ is the external force in x-direction, $\eta$ is shear viscosity, and $w$ is the pore width. We fit the velocity profiles directly obtained from NEMD simulations with a second-order polynomial

$$v_x(z) = az^2 + b, \qquad (8)$$

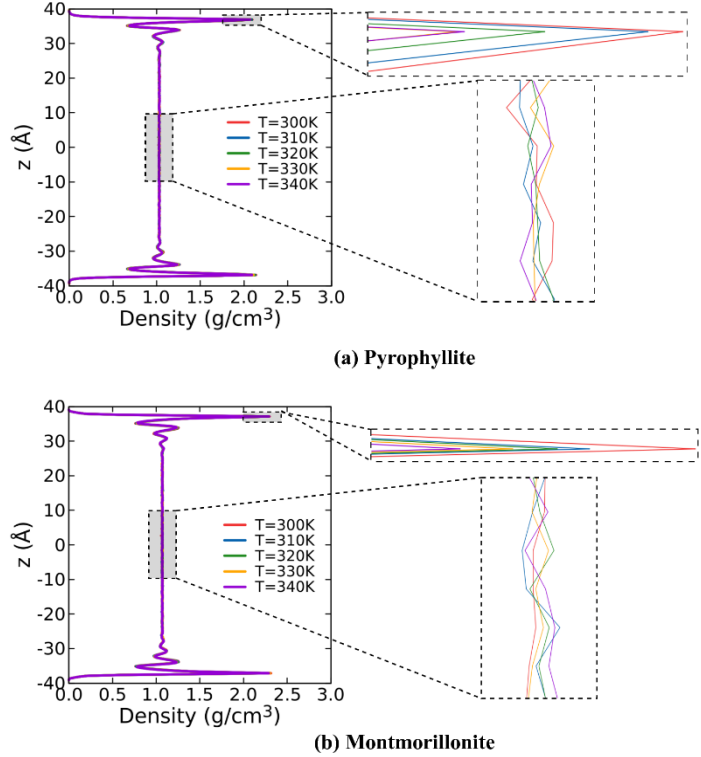

(a) Pyrophyllite

(b) Montmorillonite

Figure 6: Density profiles of confined water at different temperatures.

where $a$ and $b$ are constants. Table 3 summarizes the constants in (8) at different temperatures. From (7) and (8) the shear viscosity can be written as

$$\eta = -\frac{nF_x}{2a}, \qquad (9)$$

Table 3: Fitting constants in (8) for pyrophyllite and montmorillonite at different temperatures.

| Temperature (K) | Pyrophyllite | | Montmorillonite | |
|---|---|---|---|---|
| | a | b | a | b |
| 300 | -0.028 | 46.35 | -0.025 | 38.51 |
| 305 | -0.030 | 49.90 | -0.027 | 42.37 |
| 310 | -0.032 | 54.43 | -0.030 | 46.48 |
| 315 | -0.033 | 56.68 | -0.031 | 48.81 |
| 320 | -0.038 | 63.30 | -0.032 | 51.59 |
| 325 | -0.038 | 64.85 | -0.035 | 55.92 |
| 330 | -0.041 | 70.20 | -0.036 | 57.45 |
| 335 | -0.042 | 71.84 | -0.037 | 60.63 |
| 340 | -0.046 | 77.58 | -0.040 | 64.29 |

Figure 8 presents the parabolic velocity profiles at different temperatures. For pyrophyllite, the bulk velocity at the center of nanopore reaches 77 m/s at 340 K, 65% larger than that at 300 K. Comparing the water velocity profiles of two clay nanopores, it is found that water flow confined in pyrophyllite nanopore is approximately 19% faster than that in Na- montmorillonite in all temperature ranges. The difference in velocity can be explained by viscosity profiles between two clays. As shown in Figure 8, water viscosity witnesses a decrease of 38% as temperature decreases by 40 K. Montmorillonite nanopore produces higher viscosity than pyrophyllite due to the existence of sodium cation adsorbed by water, which enhances the shear viscosity and slows down the streaming velocity. Under the same temperature (i.e., 300 K) and using CLAYFF force field, Botan et al. [40] reported a viscosity of 0.65 ± 0.02 cP for water confined in pyrophyllite and 0.75 cP for montmorillonite. Our simulation results agree well with their data, 0.64 cP for pyrophyllite and 0.75 cP for montmorillonite.

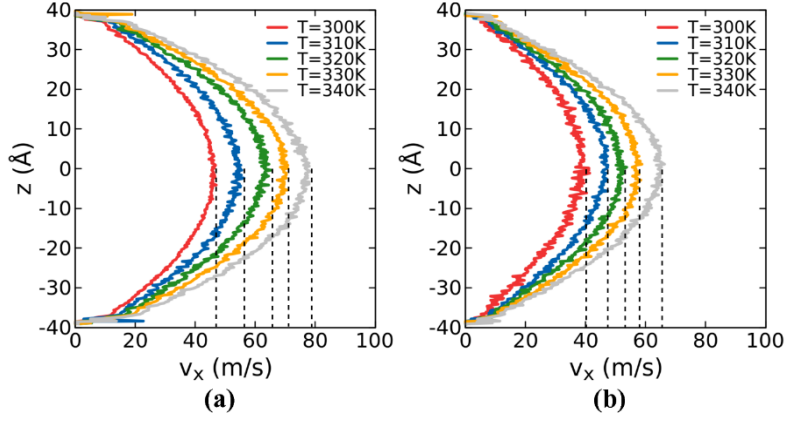

Figure 7: Effect of temperature on velocity profiles: (a) Pyrophyllite; (b) Na-montmorillonite.

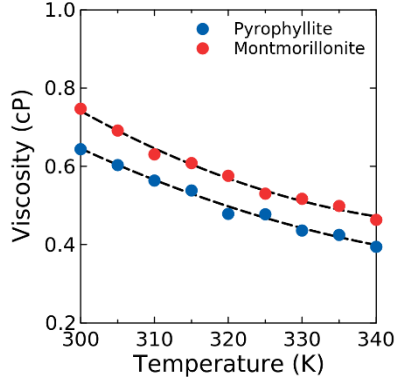

Figure 8: Effect of temperature on viscosity.

### 3.3. Clay-water slip length

In nanoscale transport phenomena, solid-fluid interactions become more significant. On the continuum scale, slip is a phenomenon describing the drag of interfacial fluid layer by adjacent layers under shear. The nonphysical quantity of slip length is usually used to characterize the slip boundary. It is defined as the distance beyond the clay surface where water velocity drops to zero. The "no-slip" or stick boundary condition in continuum hydrodynamics states that fluid at a solid-fluid interface has no relative velocity to it. However, this assumption does not hold on the atomic scale. When the fluid slips on the interface (i.e., velocity does not vanish at the interface), we adopt the concept of slip length to describe the slip boundary condition. The determination of slip length $L_s$ requires the location of clay-water interface $z_s$ as a priori. As in [40], we define $z_s$ as the location of the Gibbs dividing surface [67] that satisfies

$$\int_0^{z_s} [\rho_b - \rho(z)]dz = \int_{z_s}^{z_c} \rho(z)dz, \quad (10)$$

where $\rho_b$ is bulk water density, $\rho(z)$ are water density at $z$, and $z_c = 40$ Å is the location of clay surface. The Gibbs dividing surface is a geometrical surface that distinguishes the homogeneous bulk water from adsorbed water. In equation (10), $z_s$ can be determined using density profiles in Figure 6. Following the method in [40] $z_s = 38.125$ Å.

Slip length can be computed from the velocity profile through the velocity and its derivative at $z_s$ [14] (see Figure 10)

$$L_s = \frac{v_x}{\left|\frac{dv_x}{dz}\right|}. \quad (11)$$

Table 4: Summary of velocities and their derivatives at different temperatures at $z_s$

| Temperature (K) | Pyrophyllite | | Montmorillonite | |
| --- | --- | --- | --- | --- |
| | $v_x$ | $|dv_x/dz|$ | $v_x$ | $|dv_x/dz|$ |
| 300 | 5.07 | 2.07 | 2.31 | 1.90 |
| 310 | 9.07 | 2.37 | 3.01 | 2.28 |
| 320 | 9.81 | 2.79 | 3.93 | 2.50 |
| 330 | 11.55 | 3.06 | 4.93 | 2.74 |
| 340 | 15.58 | 3.21 | 5.85 | 2.99 |

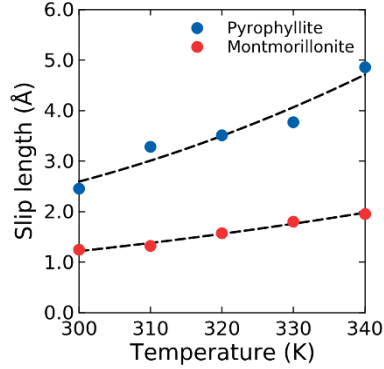

Figure 9: Effect of temperature on slip length.

Table 4 summarizes velocities and their derivatives at different temperatures at $z_s$. Figure 9 shows the temperature dependence of slip length for two clay-water systems. The barrier for water transport from one adsorption site to another could be weakened by temperature increase. Thus, the slip length becomes larger as temperature increases. It can be found from Figure 9 that montmorillonite produces a smaller slip length, which indicates that water flow confined in montmorillonite nanopore experiences a higher friction near the boundary than that in pyrophyllite nanopore. Though the physical measurement of slip length remains challenging, numerical simulations of slip clay-water boundary condition have been performed [68, 40]. For example, our NEMD simulation gives a slip length of 1.2 Å for montmorillonite at 300 K. This value is the lower bound of the range of slip length (i.e., [1.2, 1.8] Å) as reported in [69] under the same condition.

Now we can express the modified velocity equation accounting for slip length [70]

$$v_x(z) = -\frac{nF_x}{2\eta}\left(z^2 - \frac{w^2}{4} - wL_s\right). \quad (12)$$

Figure 10 compares two velocity profiles and graphically shows the slip length $L_s$. $V_{MD}$ is directly obtained from the NEMD simulation, $V_{slip}$ is the water velocity accounting for slip boundary condition from (12), and $V_{no-slip}$ is the water velocity with no-slip or stick boundary condition from (7). In the latter case water velocity drops to zero when approaching the horizontal red dash line that denotes the location of Gibbs dividing surface. The dashed line represents the location of the bottom surface of the top clay plate in Figure 2. We calculate the relative errors on velocity for slip and no-slip boundary conditions, $|V_{slip} - V_{MD}|/V_{MD} =$ 2.1% and $V_{no-slip} - V_{MD}/V_{MD} = 12.8\%$, respectively. The large relative error indicates that the applicable no-slip boundary condition at the macroscopic scale does not apply to systems at the nanoscale. Neglecting slip boundary effect results in relatively large errors in the flow velocity. Our numerical results demonstrate that the velocity equation accounting for slip boundary condition (i.e., (12)) could better describe water flow in the clay nanopore [14].

## 3.4. Hydraulic conductivity

Hydraulic conductivity is an important parameter in describing fluid flow in porous media. Its measurement in the laboratory and field scale has been well established [71]. However, the measurement technique at the nanoscale is still a challenge. In this study, we examine if cubic law [72] is applicable at the nano pore scale through the results from Darcy's law based on the numerical data. Both the cubic law and Darcy's law were proposed for laminar flow. So we need to check the Reynolds number of the water flow in the NEMD simulation setup. The Reynolds number can be expressed as

$$R_e = \frac{\rho v_x L_c}{\eta}, \qquad (13)$$

where $\rho$ is water density, $v_x$ is water velocity, $L_c$ is a characteristic length, and $\eta$ is the dynamic viscosity of water. The Reynolds number of our NEMD simulation is in the order of $10^{-3}$ that is smaller than the lower bound of the range of maximum Reynolds number (i.e., [1,10]) for laminar flow [73]. Thus the water flow through the nanoscale channel formed by two clay plates in this study can be assumed laminar flow and Darcy's law could be applied.

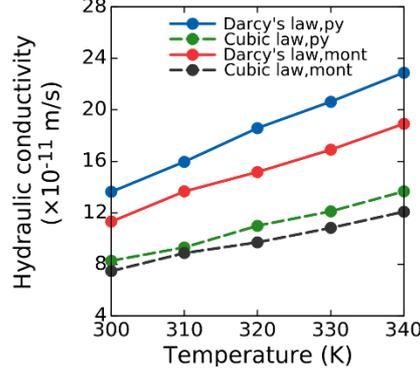
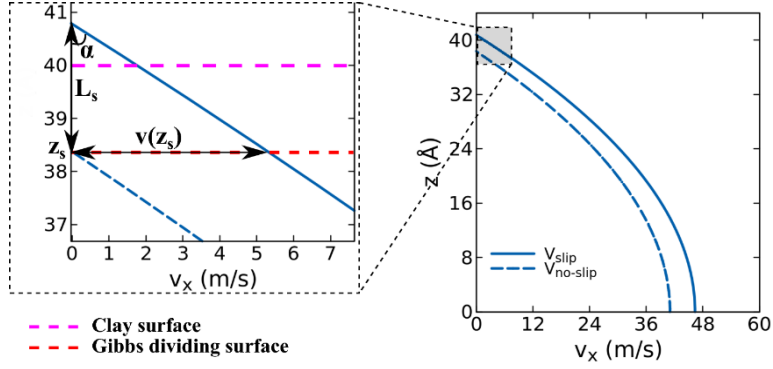

Figure 10: Comparison of two velocity profiles and the graphical representation of slip length.

Figure 11: Comparison of hydraulic conductivity calculated by cubic law and Darcy's law.

The celebrated Darcy's law describes the fluid flow through a porous medium at the continuum scale

$$Q = \frac{k}{\eta}\frac{\Delta P}{\Delta x}, \qquad (14)$$

where $Q$ is the volume flow rate, $k$ is the intrinsic permeability, $\Delta P/\Delta x$ represents the fluid pressure gradient in $x$-direction. A standard approach in describing the relationship between hydraulic conductivity $K$ and intrinsic permeability $k$ is in the following manner [74]

$$K = \frac{k\rho g}{\eta}, \qquad (15)$$

where $\rho$ is the density of confined water, and $g$ is the gravity acceleration due to gravity.

The validity of applying the cubic law to water flow in a fracture bounded by two parallel walls has been investigated by Witherspoon et al. [72]. In this case, the hydraulic conductivity is described as

$$K = \frac{w^2 \rho g}{12\eta}, \qquad (16)$$

where $w$ is the aperture of parallel-plate fracture, i.e., the width of clay nanopore. In our NEMD numerical experiment setup, water flows through the pore space formed by two clay plates that mimics water flow in a parallel-plate fracture. Thus we assume that the classical cubic law is applicable. Figure 11 compares the results of hydraulic conductivity calculated from the cubic law and Darcy's law based on the NEMD simulations in this study. With the cubic law our NEMD results yield values ranging from $7 \times 10^{-11}$ to $1.4 \times 10^{-10}$ m/s. These values are in the range of the experimental results from $2 \times 10^{-11}$ to $2 \times 10^{-9}$ m/s as reported in [75]. Figure 11 shows that the hydraulic conductivity increases almost linearly with temperature at the nanoscale. This phenomenon is consistent with experimental findings in Constantz [76]. Because the presence of sodium ($Na^+$) in the solution decreases hydraulic conductivity [77], we observe a lower value for Na-montmorillonite than pyrophyllite.

*3.5. Flow rate accounting for slip length*

To examine the effect of temperature on confined water flow, we calculate flow rate incorporating slip length. As a significant modification to the classical Navier-Stock equation used to derive volume flow rate, here we incorporate viscosity, slip length, and clay-water interface thickness. The volume flow rate along x-direction, $Q$, can now be written as

$$Q = \frac{2}{3} \frac{L_s}{\eta} \frac{\Delta P}{\Delta x} \left(\frac{w}{2} - \delta\right)^3, \qquad (17)$$

where $L_s$ is the slip length, $\Delta P$ is the pressure drop that can be determined from equation (5), $\Delta x$ is the flow length, $w$ is the pore width, and $\delta$ is the critical thickness of clay-water interface region. $\delta = 7$ Å in this study. Note that the critical thickness represents the dimension of the sticking region at the clay-water interface [78]. The similarity between critical thickness and electrical double-layer thickness is presented in the succeeding section. Figure 12 summarizes the water flow rates at various temperatures for both clay minerals in this study. Similar to the velocity-temperature relationship, water volume flow rate monotonically increases as temperature rises. It is known that clay minerals may represent very significant barriers to fluid migration. By comparison in Figure 12 we find that Na-montmorillonite yields a much lower water volume flow rate than pyrophyllite.

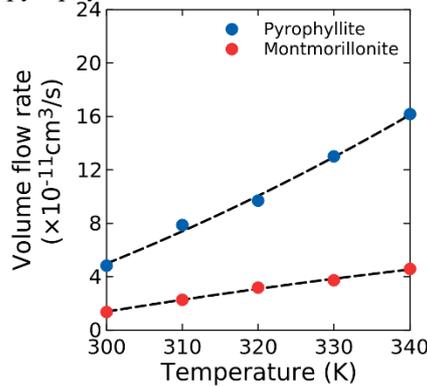

Figure 12: Water flow rate at elevated temperatures.

## 4. Discussions

In this section we discuss the adsorption behavior near the clay-water interface and calculate the clay-water friction coefficient. Potential effects of nanopore size and clay particle thickness on nanoscale hydrodynamics are also investigated.

*4.1. Adsorption behavior*

The negative charge in clay minerals is compensated by layers of interfacial water molecules. This region is the so-called electrical double layer (EDL). EDL consists of a Stern layer and a diffuse layer. Figure 13 shows the charge density profiles for water oxygen, hydrogen, and their summation denoted as "net". The charge density of an atom type is the product of its number density and atomic charge. For example, the charge density of oxygen is given by

$$\rho_e = nq_0, \qquad (18)$$

where $n$ is number density of oxygen and $q_o=-2e$ is the atomic charge of oxygen. In our simulation, the Stern layer extends from the outermost water layer to the location corresponding to the maximum magnitude of the charge density. The thickness of Stern layer for pyrophyllite and montmorillonite pore is about 1.7 Å. Water in this layer is almost immobilized [79]. The boundary of EDL is located at a position where the net charge density vanishes. Water further away from the EDL's boundary is regarded as bulk water. Thus, the thickness of EDL for pyrophyllite nanopores is 6.5 Å and the value for montmorillonite is 6.8 Å. The zero-density region close to ± 40 Å (i.e., the location of clay surface) indicates an empty water layer due to strong repulsion between clay and water.

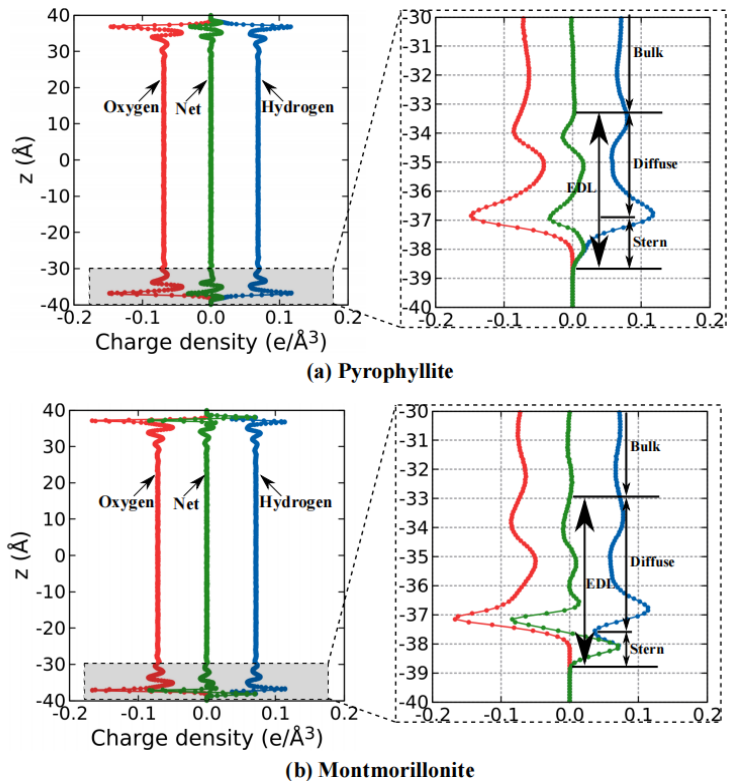

Figure 13: Schematic of electric double layer through charge density profiles (a) Pyrophyllite and (b) Montmorillonite.

Figure 14 shows the adsorption behavior of water near the clay-water interface. The results show that water hydrogens have moved closer to the clay surface than water oxygens. This observation is consistent with the number density profiles plotted in Figure 15. The first density peak adjacent to the pyrophyllite wall is 2.138 g/cm³, which is approximately 2.1 times greater than that of the bulk water. Therefore, it is reasonable to believe that this layer is in a solid-like state, indicating the strong adsorption of water onto the clay surface. To further quantify the absorption behavior of water, we identify a critical thickness $\delta$ of the clay-water interface [78]. It is defined as the one above which water is expected to behave as bulk water without being affected by solid clay. Based on water density distribution, we assume that the interface region starts from the outermost water layer to the second density peak. For the case T = 300 K, the critical thickness is about 7.0 Å for both pyrophyllite and montmorillonite. The results show that the critical thickness of clay-water interface region could be correlated to the thickness of electric double layer, 6.5 Å and 6.8 Å for pyrophyllite and montmorillonite, respectively.

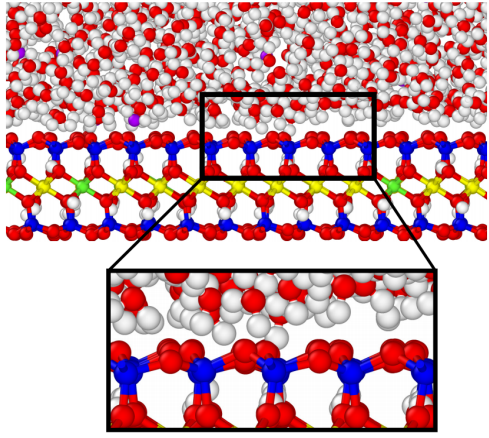

Figure 14: Water orientation near clay surface.

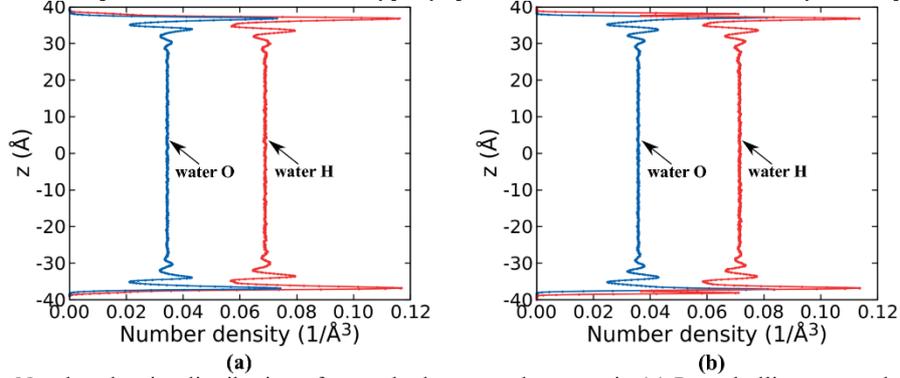

Figure 15: Number density distribution of water hydrogen and oxygen in (a) Pyrophyllite pore and (b) Montmorillonite pore.

### 4.2. Clay-water friction coefficient

In nanoscale hydrodynamics, solid-liquid friction coefficient plays an important role in flow behavior due to the presence of interfacial friction. We first introduce a friction coefficient $\lambda$ linking the tangential force per unit area $\sigma$ exerted on the clay surface to the slip velocity $v_s$ at the clay-water interface [80]

$$\sigma = n\lambda v_s. \qquad (19)$$

Combining the above equation with the constitutive equation for Newtonian water flow

$$\sigma = n\frac{dv_x}{dz}. \qquad (20)$$

one arrives at the so called Navier boundary condition [81]

$$v_s = \frac{\eta}{\lambda}\left(\frac{dv_x}{dz}\right) = L_s\frac{dv_x}{dz}. \qquad (21)$$

Then from (21) we can estimate the clay-water friction coefficient as

$$\lambda = \frac{\eta}{L_s}. \qquad (22)$$

We note that the Green-Kubo integral [14] can be used to numerically determine the friction coefficient from NEMD results. For simplicity, we determine the friction coefficient from Equation (22) in this study. Figure 16 shows the clay-water friction coefficient decreases as temperature increases. The higher friction coefficient indicates that Na-montmorillonite has a stronger ability of water adsorption. From the view of thermodynamics, interface friction explains well the mechanism of hydrodynamics variations at elevated temperature.

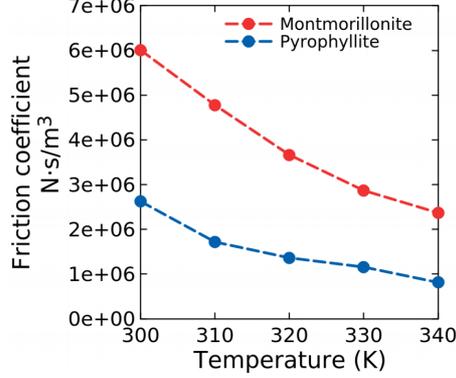

Figure 16: Clay-water friction coefficient calculated directly in NEMD and from Green-Kubo integral method at elevated temperatures.

*4.3. Effect of pore geometric size*
*4.3.1. Pore width*

To quantify the effect of pore width on hydrodynamics we conduct NEMD simulations with channel widths ranging from 8 to 14 nm, under the same temperature of 300 K and 0.0005 kcal/mol/Å driving force. Figure 17 shows the pore width effect on water velocity. The typical parabolic velocity distribution has emerged in all cases. For pyrophyllite, as pore width increases from 8 to 14 nm, flow velocity at the central pore increases from 45.7 to 143.0 m/s. By comparison, the montmorillonite pore produces lower water flow velocity, from 38.1 to 127.9 m/s, at the same pore width.

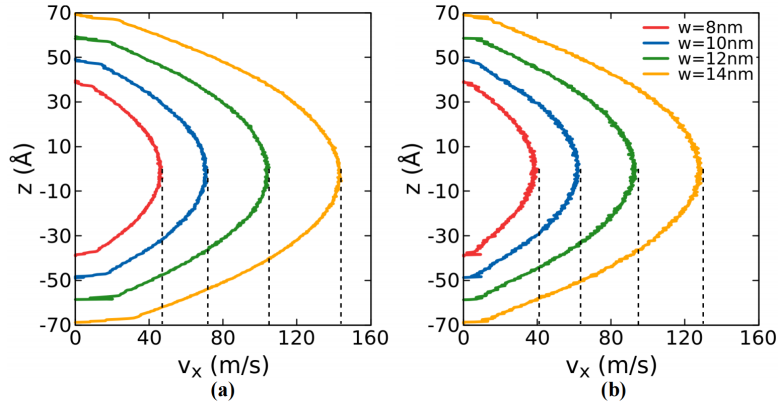

Figure 17: Effect of pore width on water velocity (a) pyrophyllite, (b) montmorillonite.

Figure 18 shows the pore width effect on water density. The density peak near clay surface can be found in all widths. However, the peak value slightly decreases with the increase of pore width. Comparison of two types of clay pores show that montmorillonite pore gives larger peak density (approximately 0.25 g/cm$^3$) than pyrophyllite.

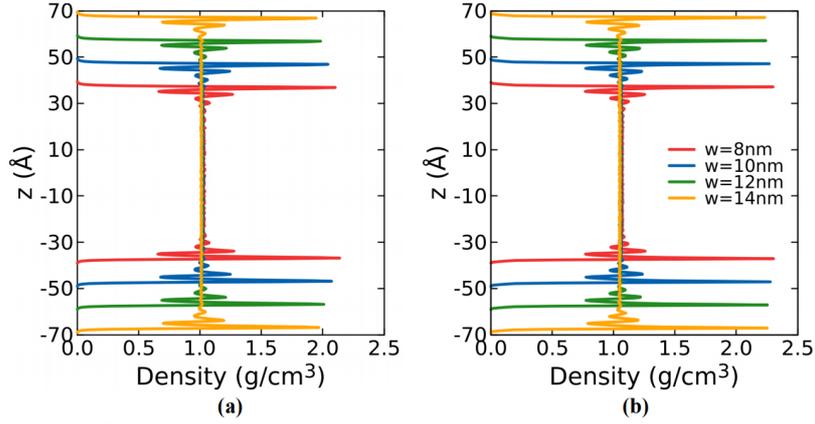

Figure 18: Effect of pore width on water density (a) pyrophyllite, (b) montmorillonite.

Figures 19 and 20 present the effect of pore width on water viscosity and slip length, respectively. Minor changes occur in viscosity while a monotonous increase in slip length with wider pore can be observed. The relationship between clay-water slip length and pore width agrees well with results in Sokhan and Quirke [82]. Montmorillonite pore yields larger viscosity and smaller slip length than pyrophyllite. The difference above proves that montmorillonite has a stronger capacity to attract and re-orientate neighboring water molecules than pyrophyllite due to sodium cations.

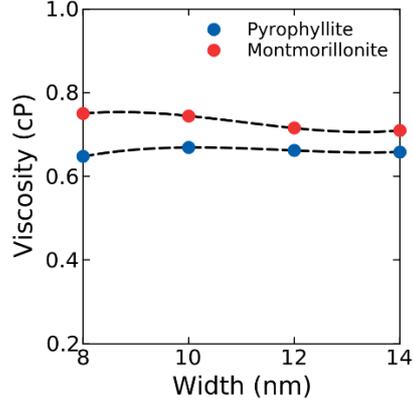

Figure 19: Effect of pore width on water viscosity.

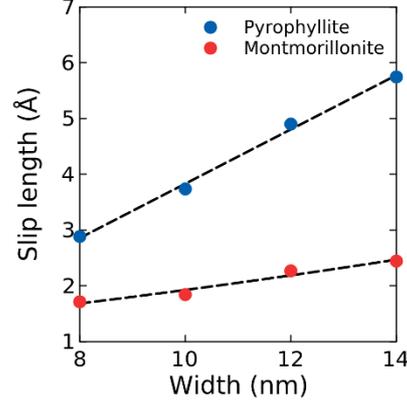

Figure 20: Effect of pore width on slip length.

### 4.3.2. Thickness of clay particle

We investigate the effect of clay particle thickness in terms of the number of repeating layers in z-direction on hydrodynamic behaviors. Since a cut-off value of 10 Å is used in this study, we can reasonably assume that the contribution from the van der Waals forces is negligible. However, the long-range electrostatic interactions may affect hydrodynamics with increased particle thickness. Thus, we perform two groups of simulations, one employing a single layer and the other employing two layers to model the clay particle.

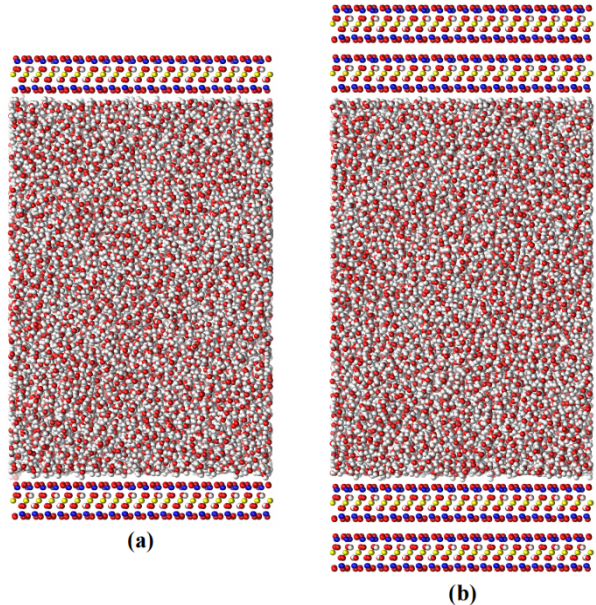
Figure 21: Clay particles with different thicknesses.

Table 5: Comparison of hydrodynamics between 1- and 2-layer clay particles

| Particle | Peak density (g/cm$^3$) | Velocity (m/s) | Viscosity (cP) | Hydraulic conductivity (m/s) |
|---|---|---|---|---|
| 1-layer | 2.331 | 48.82 | 0.66 | $15.0\times10^{-11}$ |
| 2-layer | 2.336 | 49.47 | 0.67 | $15.2\times10^{-11}$ |
| Deviation (%) | 0.2 | 1.3 | 1.5 | 1.3 |

The results show that the 2-layer structure has a weak effect on hydrodynamics compared to 1-layer structure. Deviations are less than 1.5%. This reflects the long-range electrostatic interactions decay as the atomic distance becomes larger. Thus, almost no effect is found on hydrodynamic properties.

## 5. Closure

We have investigated hydrodynamics of water flow confined in clay nanopores at elevated temperatures via NEMD simulations. The system under investigation is water confined in a nanochannel formed by two parallel clay particles (i.e., pyrophyllite and montmorillonite). We studied the temperature effect on hydrodynamic properties including water density, flow velocity, shear viscosity, and hydraulic conductivity accounting for slip clay-water boundary condition. Based on our simulation data, we examined the temperature effect on water flow velocity, shear viscosity, clay-water slip length, hydraulic conductivity, and clay-water friction coefficient at the nanopore scale. For instance, as temperature rises by 40 K, the viscosity drops by about 40%, and pyrophyllite-water slip length increases by 200%. Our numerical results showed a monotonically increasing relationship between hydraulic conductivity and temperature. Increasing the temperature by 40 K results in a decrease of 54% in the clay-water friction coefficient. We also investigated the impact of pore size and clay structure on hydrodynamics. When the pore size becomes larger, the flow is enhanced with more incredible streaming velocity and longer slip length. The number of clay layers has a mild effect on water flow velocity, viscosity, and hydraulic conductivity. Our work has shown that NEMD is a promising tool in evaluating the nanoscale hydrodynamic proprieties of clay at elevated temperature. The hydrodynamic properties from NEMD simulations in this study could be useful for developing a physics-based multiscale model for fluid transport in clays through homogenization/upscaling techniques (e.g., [83, 12, 11]) that is an ongoing effort.

## Acknowledgments

This work has been supported by the US National Science Foundation under contract numbers 1659932 and 1944009. The authors are grateful to the two anonymous reviewers for their constructive reviews of the first version of this article.

**Data Availability Statement**

The data that support the findings of this study are available from the corresponding author, X. S., upon reasonable request.